\documentclass[ twocolumn,preprintnumbers ,amsmath, amssymb, superscriptaddress, aps]{revtex4}

\usepackage{color}
\usepackage{amsmath}
\usepackage{pifont}
\usepackage{amssymb}  
\usepackage{bbold}
\usepackage{float}
\usepackage{tikz}
\usepackage{makecell}
\usepackage{subfigure}
\usepackage{pifont}   
\usepackage{graphicx} 
\usepackage{dcolumn}  
\usepackage{bm}       
\usepackage{multirow} 
\usepackage{hyperref}
\usepackage{placeins}
\newcommand{\vect}[1]{\mathbf{#1}}

\newcommand\vex[1]{\mathbf{#1}}

\begin{document}

\renewcommand\floatpagefraction{0.8} 
\renewcommand\topfraction{0.8}       

\author{Michael Vogl}
\affiliation{Department of Physics, The University of Texas at Austin, Austin, TX 78712, USA}
\author{Martin Rodriguez-Vega}
\affiliation{Department of Physics, The University of Texas at Austin, Austin, TX 78712, USA}
\affiliation{Department of Physics, Northeastern University, Boston, MA 02115, USA}
\author{Gregory A. Fiete}
\affiliation{Department of Physics, Northeastern University, Boston, MA 02115, USA}
\affiliation{Department of Physics, Massachusetts Institute of Technology, Cambridge, MA 02139, USA}
\title{Effective Floquet Hamiltonian in the low-frequency regime}
\date{\today}

\begin{abstract}
We develop a theory to derive effective Floquet Hamiltonians in the weak drive and low-frequency regime. We construct the theory in analogy with band theory for electrons in a spatially-periodic and weak potential, such as occurs in some crystalline materials. As a prototypical example, we apply this theory to graphene driven by circularly polarized light of low intensity. We find an analytic expression for the effective Floquet Hamiltonian in the low-frequency regime which accurately predicts the quasienergy spectrum and the Floquet states. Furthermore, we identify \textit{self-consistency} as the crucial feature effective Hamiltonians in this regime need to satisfy to achieve a high accuracy. 
 The method is useful in providing a realistic description of off-resonant drives for multi-band solid state systems where light-induced topological band structure changes are sought.
\end{abstract}
\maketitle

\textit{Introduction.} Recent years have seen rapid developments in our understanding of systems out of equilibrium. Experimental advances in ultra-fast spectroscopy have lead to the observation of Floquet side bands~\cite{wang2013,huang2018}, the discovery of light-induced superconductivity~\cite{fausti2011,mitrano2016}, and light-induced anomalous Hall effect in graphene~\cite{mcIver2018}, just to name a few of the most striking examples. On the theory side, efforts have lead to the prediction of new phases of matter without equilibrium counterparts--and in some cases subsequent experimental observations. Examples are  Floquet time crystals ~\cite{Dominic2016,PhysRevLett.118.030401,PhysRevX.7.011026,10.1038/nature21426,10.1038/nature21413}, and anomalous Anderson insulators~\cite{titum2016}. Additionally, the potential \textit{in situ} manipulation of topological phases by tuning the properties of the drive has lead to plethora of motivating predictions~\cite{oka2009,PhysRevX.3.031005,lindner2011,tong2013,PhysRevB.88.155133, kundu2013,rechtsman2013x,jiang2011,zhenghao2011,PhysRevB.89.121401,PhysRevB.90.115423,PhysRevA.91.043625,PhysRevB.91.241404,PhysRevB.97.205415,PhysRevB.97.245401,2019arXiv190902008R,PhysRevB.91.155422,PhysRevB.92.165111,PhysRevB.93.205437}.

In a periodically-driven system, we can distinguish three regimes as a function of the drive frequency $\Omega$ compared with the bandwidth of the system $W$: the high, mid (or resonant), and low-frequency regimes. In the high-frequency regime $\Omega > W$, several theoretical approaches have been developed and are now widely applied in the field to derive effective Floquet Hamiltonians capable of capturing the dominant effects of the periodic drives~\cite{PhysRevB.95.014112,blanes2009, bukov2015, eckardt2015,mikami2016,bukov2016}. This is highly desirable since it allows one to employ many equilibrium techniques to study systems of interest. In the resonant and low-frequency regimes, $\Omega \lesssim W$, the high-frequency expansions break down and one generally needs to resort to numerical approaches. More recently, efforts to understand the mid and low-frequency regimes have led to the use of rotating frames \cite{2019arXiv190904064H,2018arXiv180304490L}, the development of adiabatic perturbation theories~\cite{Martiskainen2015,rigolin2008}, low-frequency perturbation theories in the extended Floquet-Hilbert space~\cite{weinberg2015,Jia-Ming2016,Rodriguez_Vega_2018}, and renormalization group-like flow equation schemes~\cite{verdeny2013,vogl2019}. 
Despite this progress, a systematic theory to derive effective Hamiltonians valid in the low-frequency regime is still missing. 

\begin{figure}[t]
	\centering
	\includegraphics[width=1\linewidth]{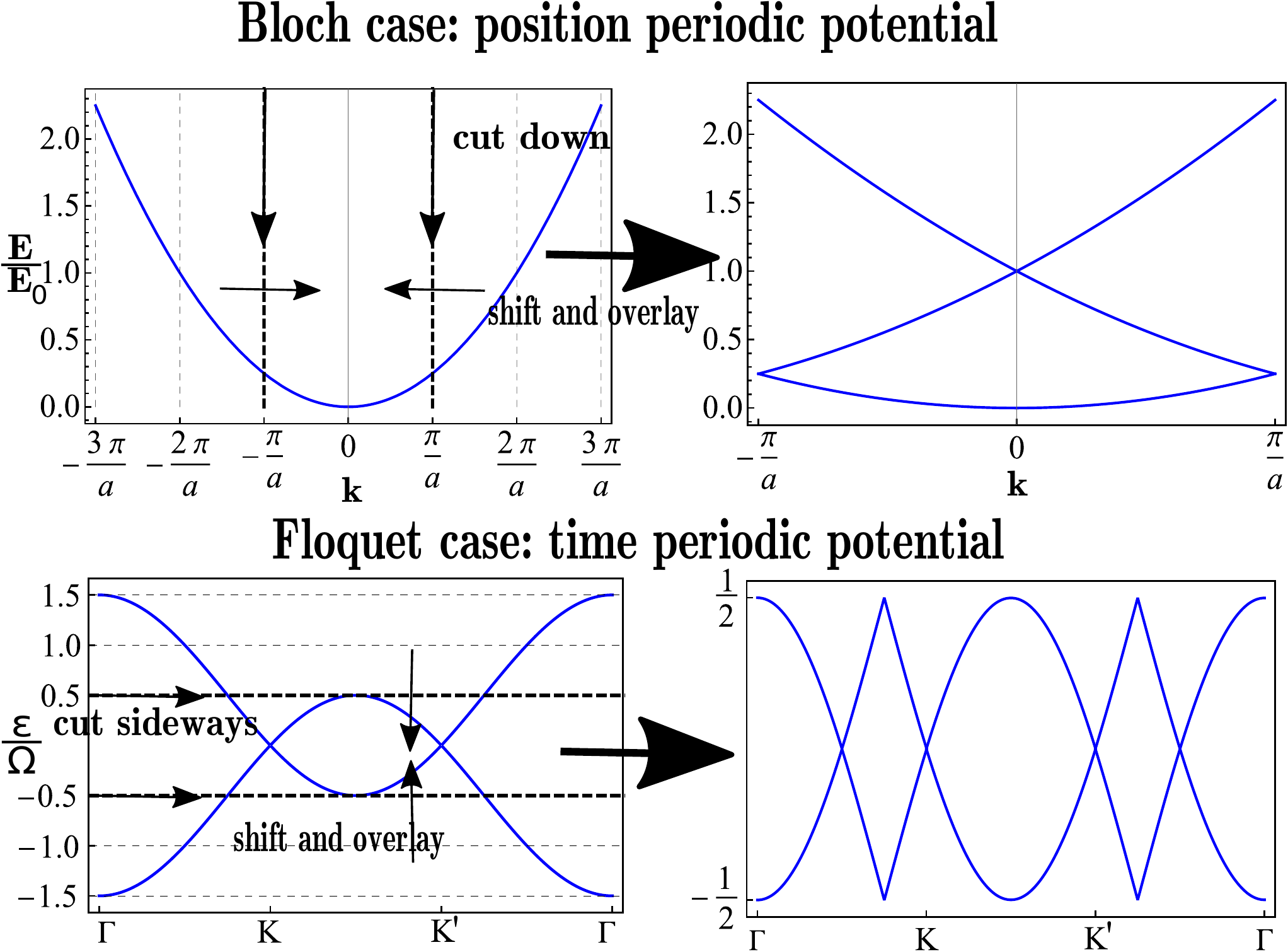}
	\caption{(Color online.) Visual schematic of zone folding scheme.  In the upper row we see a schematic of how the different Brillouin zones are cut apart vertically and all shifted to the first Brillouin zone. In the lower row we see how the band structure of graphene needs to be cut along the horizontal direction to find the zeroth approximation Floquet bands in the first Floquet zone. }
	\label{fig:drawing-foldingstuff}
\end{figure}

In this paper, we derive effective Floquet Hamiltonians in the low-frequency regime characterized by $\Omega \lesssim W$. This regime is relevant for many experiments, since the use of low-frequency and low-power drives can reduce unwanted heating effects in interacting systems. Additionally, it has been predicted that this regime hosts intriguing phenomena unique of this regime~\cite{Rodriguez-Vega2018b,privitera2018,kim2019}. 

Our approach is general, and allows one to obtain analytic insight into periodically driven systems and provides a quantitative description. In analogy with a system periodic in space, we first employ an \textit{empty lattice-type approximation} to gain insight into the possible effects of weak periodic drives in the first Floquet zone $(-\Omega/2, \Omega/2]$. Resonances at the Floquet zone center ($\epsilon/\Omega = 0$) and edge ($\epsilon/\Omega = 1/2$) can be resolved by employing a continued fraction approximation scheme~\cite{2015PhRvA..91c3416P,10.1088/2515-7639/ab387b}.   Here $\epsilon$ is the quasi-energy.

\textit{Model}. Specifically, to demonstrate our method, we consider graphene weakly driven with circularly polarized light, with a time-dependent Hamiltonian of the form $\mathcal H(t) = \int_{BZ} d \vex k/(2\pi)^2  \hat c^\dagger_{\vex k} h(\vex k, t) \hat c_{\vex k} $ where the integration over crystal momentum is defined over the Brillouin zone (BZ),  $\hat c^\dagger_{\vex k} $ is the creation operator, and
\begin{equation}
	h(\vex k, t)  =\begin{pmatrix}	0&f(\vex k,t)\\
	f^*(\vex k,t)&0
	\end{pmatrix},
\end{equation}
$f(\vex k,t)=e^{\frac{i \tilde k_x (t)}{2}-\frac{i}{2}  \sqrt{3} \tilde k_y (t)}+e^{\frac{i \tilde k_x(t)}{2}+\frac{ i}{2} \sqrt{3} \tilde k_y(t)}+e^{-i \tilde k_x(t)}$ is the kernel in the plane-wave basis. Circularly polarized light with field strength $A$ is introduced via minimal substitution as $\tilde k_x(t) = k_x - A \cos (\Omega t)$, and $\tilde k_y(t) = k_y - A \sin (\Omega t)$ where we work in natural units $\hbar = c = e = 1$. The exact dynamics can be obtained by solving the Floquet-Schr\"odinger equation $[h(\vex k, t)-i\partial_t] \phi^\pm(\vex k, t) = \pm\epsilon (\vex k) \phi^\pm (\vex k, t)$ for the steady-states $\phi^\pm(\vex k, t) = \phi^\pm(\vex k, t+2\pi/\Omega)$, and quasienergy $\pm \epsilon(\vex k)$  in the first Floquet zone $(-\Omega/2,\Omega/2]$, using the Floquet evolution operator $ U(\vex k, 2\pi/\Omega) = \mathcal T \exp\{-i \int^{2\pi/\Omega}_0  h(\vex k, s) ds\}$. 

Alternatively, we can exploit the periodicity of the steady states to define the Fourier series $\phi^\pm(\vex k, t) = \sum_n e^{i n \Omega t} \phi^\pm_n(\vex k)$, where $\phi^\pm_n(\vex k)$ are the steady-state Fourier modes. Replacing this in the Floquet-Schr\"odinger equation, we obtain the equation $\sum_m \left(h^{(n-m)} - m\delta_{n,m} \Omega \right) \phi^\pm_m(\vex k) = \pm\epsilon (\vex k) \phi^\pm_n(\vex k)$, defined in the extended Floquet Hilbert space $\mathfrak F = \mathfrak H \otimes \mathfrak I$, where $\mathfrak H$ is the Hilbert space for $h(\vex k, t)$, and  $\mathfrak I$ is spanned by a set of bounded periodic function defined over the interval $t \in [0, 2\pi/\Omega)$.

\textit{Time analog of the empty lattice picture}. To understand how weak periodic drives of arbitrary frequency impact quantum systems let us first recall the analogous spatially periodic case. For a Hamiltonian that is time-independent and periodic in space, $\mathcal H(\vect r)= \mathcal H(\vect r+\vect R)$, it is commonly observed \cite{kittel2004introduction,ashcroft1976solid,marder2010condensed} that even weak periodic potentials lead to the complicated collection of band structures common in solid state physics. If we treat interactions with the lattice as infinitesimally weak, the only effect of a periodic potential is that momentum space breaks up into periodically repeating sections called Brillouin zones (BZs). 

The BZ have shapes that are determined by the lattice geometry in real space. In each BZ one has repeated a copy of the energy band of a free electron. These different copies of the bands from various BZ overlap. The shape of the BZs then determine where different copies of the electron bands start and therefore in which way they overlap. This ultimately leads to a complicated collection of band structures that is determined by the spatial geometry. 

Instead of overlapping different copies of the electron band structure from different BZ, the same effect is produced if one takes only one free electron band centered in the first Brillouin zone--and no copies in the others--and then moves the content of the other BZ into the first zone similar to what is shown in Fig.\ref{fig:drawing-foldingstuff}. While this approximation--dubbed the empty lattice approximation because the spatially periodic potential is set to zero and only symmetry properties are kept--is quite crude, it is a good first-order estimate. Indeed, aluminum is a material where the empty lattice approximation reproduces band structures quite well (see Ref.[\onlinecite{PhysRev.118.1182}] for a band structure that can be compared to the empty lattice approximation result). 

For the time-periodic Hamiltonians, an analogous situation occurs. One finds multiple copies of the band structure along the quasienergy axis. These could overlap if the drive frequency $\Omega$ is low enough. Similar to the spatially periodic case, reproducing the effects of multiple intersecting Floquet zones can be done easily.  It is sufficient to apply a shifted modulo function, defined as $\mathrm{Mod}(E,\Omega,-\Omega/2)$,  to the energies to find the spectrum in the first Floquet zone. The effects of both cases, spatially and time periodic, are displayed in Fig. \ref{fig:drawing-foldingstuff}. One can clearly observe the development of more complicated structures in both cases compared with the uniform cases. It is also important to notice that quite generically this type of folding leads to band-crossings. Only specific properties of a periodic perturbation can lift the band crossings.

\begin{figure}[t]
	\centering
	\includegraphics[width=0.9\linewidth]{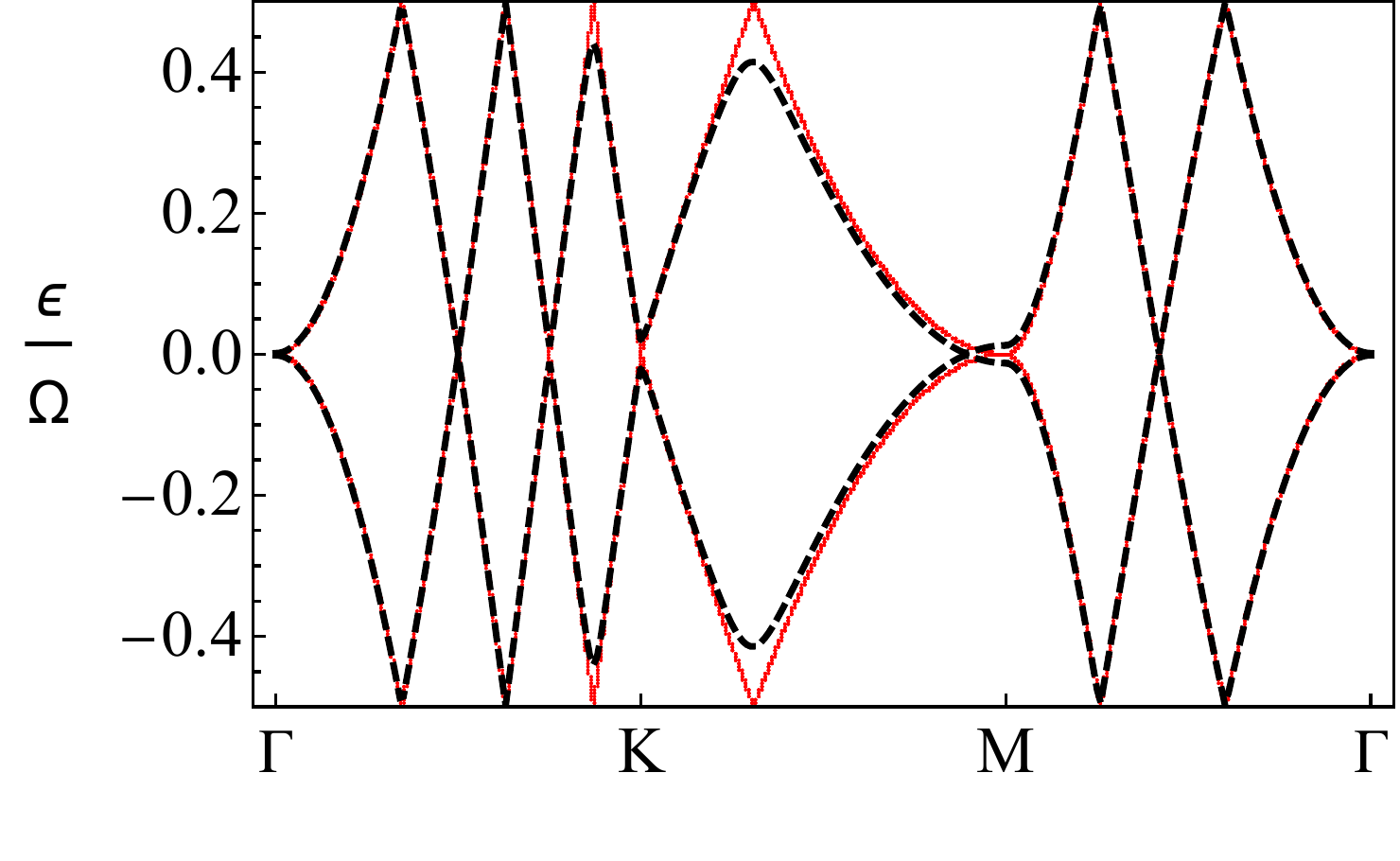}
	\caption{(Color online) Plot of the Floquet quasi-energy band structure along a path between high symmetry points for graphene at $\Omega=1$ and $A=0.1$. Black dashed we see the exact solution found by diagonalizing $i\log (U(\vex k,T))/T$ numerically, where $U(\vex k, T)$ is the propagator at time $T$. In red we see the the zero driving Floquet approximation.}
	\label{fig:undrivenfloquet}
\end{figure}

Let us now analyze how well this approximation captures the quasi-energy band structure for graphene driven by circularly polarized light. In Fig.\ref{fig:undrivenfloquet} we compare the empty-lattice-type approximation to the exact quasi-energy bands. One finds that the results are accurate in many parts of the Brillouin zone, but important band gap openings are not reproduced. It is the subject of the rest of this work to study how to obtain the correct bandgap openings due time-reversal symmetry breaking in graphene irradiated with circularly polarized light.


\textit{Quasienergies}. For a monochromatic drive, or for a general drive in the weak-drive limit, characterized by $A \ll 1$,
 the time-dependent Hamiltonian takes the general form
 \begin{equation}
 h(t)=  h_0+\mathcal Pe^{-i\Omega t}+\mathcal P^\dag e^{i\Omega t},
 \label{eq:monocromatic}
 \end{equation}
 where $h_0$ is the static Hamiltonian, and $\mathcal P$ the first-harmonic operator. 
In the extended Floquet-Hilbert (Sambe) space, the Floquet-Schr\"odinger equation can be written as~\cite{2015PhRvA..91c3416P,10.1088/2515-7639/ab387b,PhysRevA.7.2203,Eckardt_2015,PhysRevA.95.023615,PhysRevLett.111.175301,Rodriguez_Vega_2018} 
 \begin{align}
 	&\begin{pmatrix}
 	\ddots&\vdots&\vdots&\vdots&\vdots&\vdots&\\
 	\cdots&P^\dag& h_0-\Omega&P&0&0&\cdots\\
 	\cdots&0&P^\dag&  h_0&P&0&\cdots\\
 	\cdots&0&0&P^\dag& h_0+\Omega&P&\cdots\\
 	&\vdots&\vdots&\vdots&\vdots&\vdots& \ddots
 	\end{pmatrix}\begin{pmatrix}
 	\vdots\\
 	\phi_{-1}\\
 	\phi_0\\
 	\phi_1\\
 	\vdots
 	\end{pmatrix}\\&=\epsilon \begin{pmatrix}
 	\vdots\\
 	\phi_{-1}\\
 	\phi_0\\
 	\phi_1\\
 	\vdots
 	\end{pmatrix}.
 	\label{quasi-en-eq}
 \end{align}

This equation can be decoupled \cite{2015PhRvA..91c3416P,10.1088/2515-7639/ab387b} into an equation for the first Floquet mode $\phi_0$ only. The result is the continued fraction
 \begin{equation}
 \begin{aligned}
&h_{\mathrm{eff}}=h_0+P\frac{1}{\epsilon-h_0-\Omega-P\frac{1}{\epsilon-h_0-2\Omega-\cdots}P^\dag}P^\dag\\
&+P^\dag\frac{1}{\epsilon-h_0+\Omega-P^\dag\frac{1}{\epsilon-h_0+2\Omega-\cdots} P} P.
 \end{aligned}
 \end{equation}

 For weak driving, the continued fraction can be truncated to linear order in $P$, such that 
 \begin{equation}
 	h_{\mathrm{eff}}\approx h_0+P\frac{1}{\epsilon-h_0-\Omega }P^\dag+P^\dag\frac{1}{\epsilon-h_0+\Omega} P.
 	\label{truncfrac}
 \end{equation}
 
For graphene driven by a weak field, defined by $A \ll 1$ ($a_0 eA/\hbar \ll 1$ in physical units) of circularly polarized light, the Hamiltonian to first order in field strength $A$ is monochromatic with the same structure as Eq.\eqref{eq:monocromatic}. We find that
\begin{equation}
	\begin{aligned}
	&P^\dag=\begin{pmatrix}
	0&p_+\\
	p_-&0
	\end{pmatrix},\\
	&p_\pm  =\pm\frac{1}{2} i  e^{\mp i k_x } \left(1\pm 2  e^{\pm\frac{3 i k_x }{2}} \sin \left(\frac{1}{6} \left(3 \sqrt{3} k_y\mp\pi  \right)\right)\right).
	\end{aligned}
\end{equation}
Therefore, Eq.\eqref{truncfrac} can be used to find an effective energy dependent Hamiltonian $h_{\mathrm{eff}}(\epsilon)$ that reproduces the quasienergy spectrum in the first Floquet zone,
\begin{equation}
 h_{\mathrm{eff}}(\epsilon)=  h_0+A^2 \left(\mathcal M_++\mathcal M_- \right),
\label{eff_graphene}
\end{equation}
where
\begin{align}
\mathcal M_{\pm} & =\frac{1}{(\epsilon\pm\Omega)^2-|f|^2}\begin{pmatrix}
|p_\mp|^2(\epsilon\pm\Omega)&f^*p_-^*p_+\\fp_-p_+^*&|p_\pm|^2(\epsilon \pm\Omega).
\end{pmatrix}
\end{align}

Equation~(\ref{eff_graphene}) is the main result of this work, which corresponds to an analytic expression for the effective Floquet Hamiltonian valid in the low-frequency limit.

\begin{figure}[t]
	\centering
	\includegraphics[width=1\linewidth]{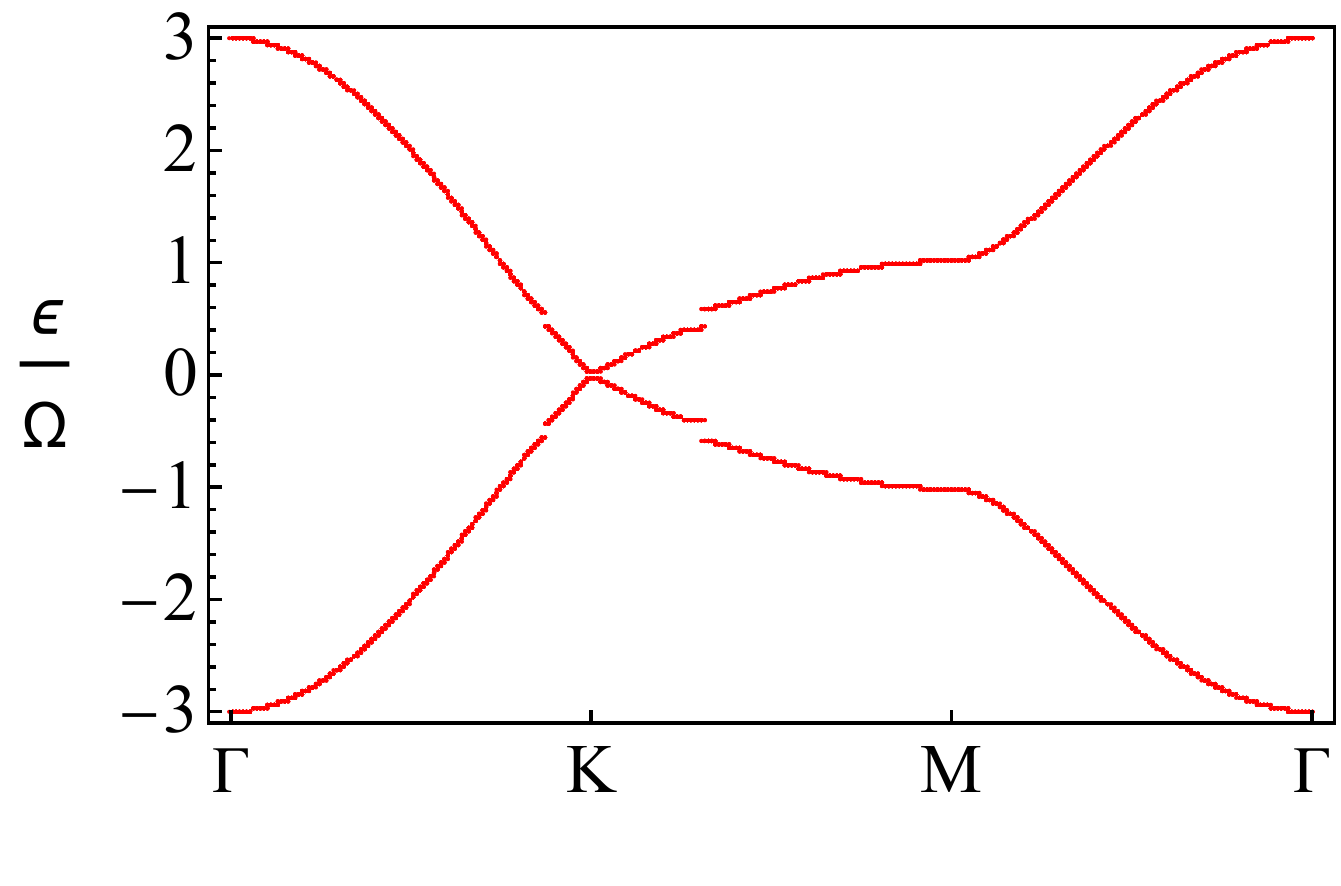}
	\caption{(Color online.) Plot of the Floquet quasi-energy band structure along a path between high symmetry points for graphene at $\Omega=1$ and $A=0.1$. Quasi energies are outside the first Floquet zone and describe only one Floquet copy.}
	\label{fig:placeholder}
\end{figure}

In order to determine the quasi-energy spectrum and steady-state mode $\phi_0$, we self-consistently solve the Schr\"odinger eigenvalue equation $\left( h_{\mathrm{eff}}(\epsilon) - \epsilon \right) \phi_0=0$. We remark that by self-consistency we mean that the Schr\"odinger equation above is solved for a fixed value of $\epsilon$ and new eigenvalues $\epsilon$ are found that are then reinserted into the Hamiltonian as fixed $\epsilon$. This procedure is repeated until self-consistency for $\epsilon$ is reached. The self-consistency is crucial, since a first order iterative approach yields divergences and is therefore unphysical. The results from such a self-consistent solution are shown in Fig.~(\ref{fig:placeholder}). 

\begin{figure}[t]
 	\centering
 	\includegraphics[width=1\linewidth]{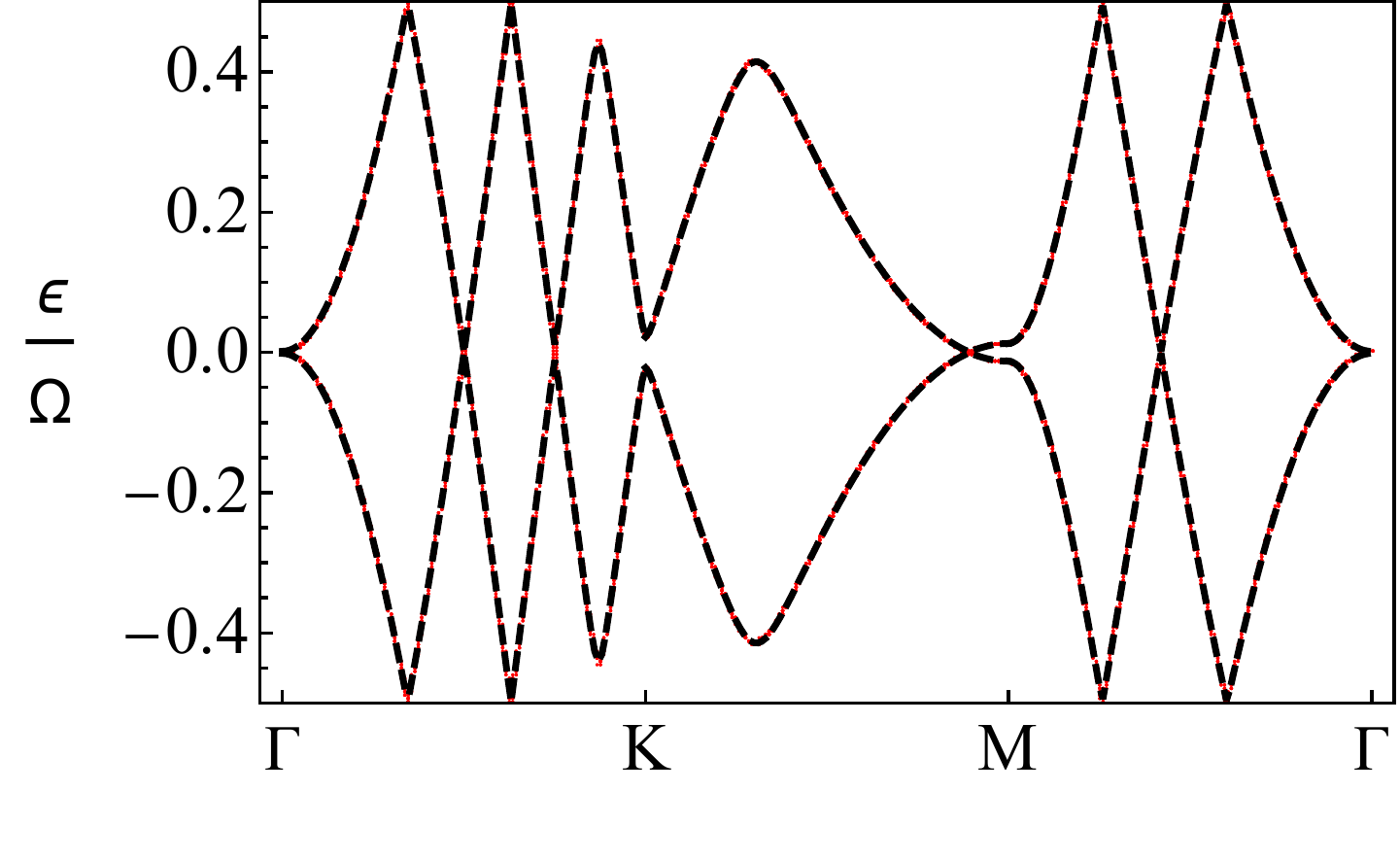}
 	\caption{(Color online.) Plot of the Floquet quasi-energy band structure along a path between high symmetry points for graphene at $\Omega=1$ and $A=0.1$. Dashed in black we plotted the numerically exact result that is found by diagonalizing $i\log (U(\vex k, T))/T$, where $U(\vex k,T)$ is the propagator at time $T$.In red we plotted our approximate result.}
 	\label{fig:almost-undriven-floquet}
 \end{figure}

 We first notice that the band structure has discontinuities that correspond to the bandgap opening at the edges of the first Floquet zone. This becomes clear when we apply the modulo function to fold the quasienergy bands into the first Floquet zone, as shown in Fig.~\ref{fig:almost-undriven-floquet}. We stress that the folding procedure used to arrive at this plot is important especially for when we later calculate Chern numbers - simply solving for $\phi_0$ with the Hamiltonian \eqref{truncfrac} is not enough. At the $\Gamma$ point, the quasienergies are given by $\epsilon= \text{Mod}\left[\pm3,\Omega ,-\frac{\Omega }{2}\right]$. Then, $\Gamma$-point gap closings at the Floquet zone center and edge occur at drive frequencies $\Omega=6/(2n)$ and $\Omega=6/(2n+1)$ respectively, \textit{independent} of the details of the drive. At the $K$ (and also $K'$, by symmetry), the quasienergy bandgap is given by $\Delta_K=\sqrt{9A^2+\Omega^2}-\Omega$. This result is non-perturbative, and even correctly gives the high frequency result  $\Delta_K=9A^2/(2\Omega)$.

A few remarks are in order. This exercise allowed us to identify key ingredients an effective Hamiltonian needs to accurately describe the low frequency regime. Importantly, such a Hamiltonian necessarily needs to be quasienergy dependent because this is what allows the occurrence of discontinuities in the quasienergy band. These are needed because they can lead to bandgap openings at the edges of the first Floquet zone--otherwise applying $\text{Mod}\left[\epsilon,\Omega ,-\frac{\Omega }{2}\right]$ to $\epsilon$ would lead to cusps rather than bandgaps. These types of jumps are non-analytic in momentum and therefore cannot be captured by finite range tight binding models. We therefore cannot expect the Magnus expansion or similar expansions to capture this type of behavior at \textit{any} finite order.

\begin{figure}[H]
	\centering
	\includegraphics[width=1\linewidth]{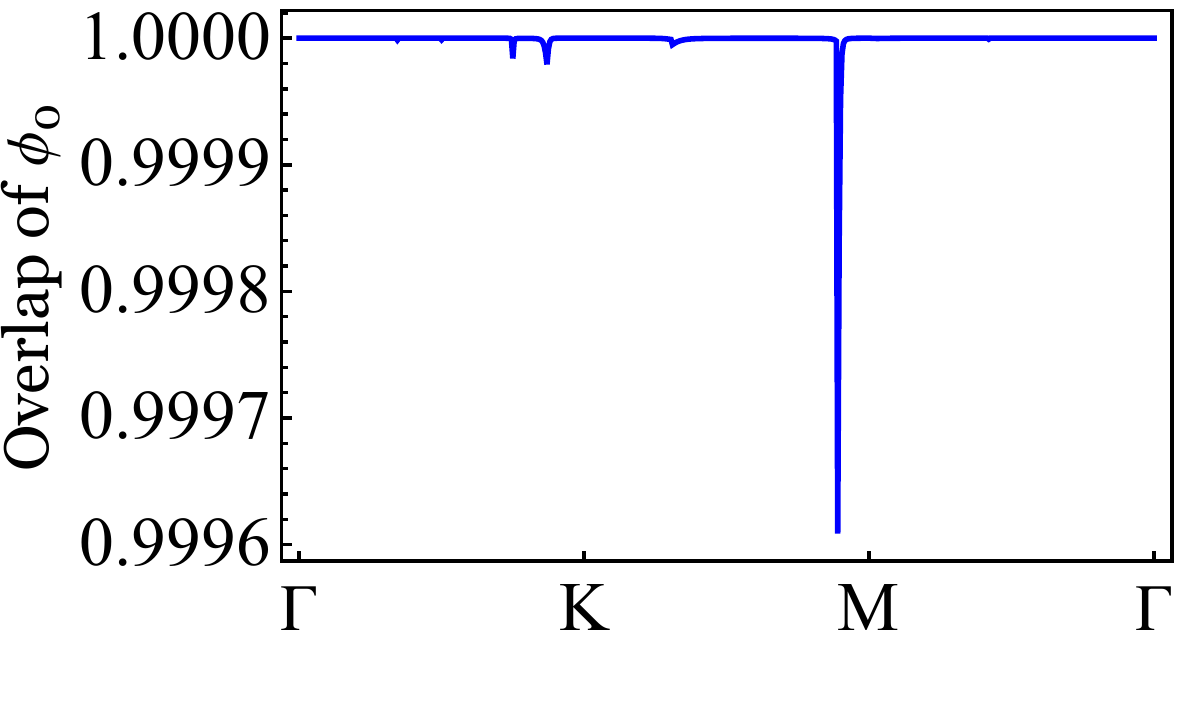}
	\caption{(Color online.) Overlap between the numeric solution of $\phi_0$ for Eq.\eqref{quasi-en-eq} with $\Omega=1$ and $A=0.1$ solved for 20 Floquet modes and our approximate result plotted along a high symmetry point in the Brillouin zone.}
	\label{fig:u0overlap}
\end{figure}

\textit{Wavefunctions.} When studying topological properties of Floquet systems, it is crucial to find an accurate approximation for all the components $\phi_n$ of a Floquet steady-states $\phi$. Now, we outline the procedure to build the steady-states from $\phi_0$. Following Ref.[\onlinecite{2015PhRvA..91c3416P,10.1088/2515-7639/ab387b}] this can be done by making use of the recursion relation
\begin{equation}
(\epsilon+m\Omega-h_0)\phi_m=P^\dag \phi_{m-1}+P\phi_{m+1}.
\label{recursion}
\end{equation}
All components of $\phi$ can thus be obtained from $\phi_0$. Therefore it is sufficient to check if $\phi_0$ is well approximated over the Brillouin zone. This is done in Fig.\ref{fig:u0overlap}, where Eq.\eqref{quasi-en-eq} is numerically solved for 20 Floquet modes.

The overlap is almost unity across the whole BZ. It is useful to realize that Eq.\eqref{recursion} can for weak driving i.e. small $P$ be approximated as
\begin{equation}
\begin{aligned}
&\phi_{n}=\frac{1}{\epsilon+m\Omega-h_0}P^\dag \phi_{n-1}\\ &\phi_{-n}=\frac{1}{\epsilon-m\Omega-h_0}P \phi_{-n+1}
\end{aligned},
\end{equation}
which is consistent with the rest of our approximation and which we use in the following section to find $\phi$.

Below we plot the overlap between $\phi_0$ (for \eqref{quasi-en-eq} truncated at 16 Fourier modes) and its approximate version. The results were averaged over a the high-symmetry path $\Gamma\to K\to M\to \Gamma$ to be able to plot them as a function of driving frequency $\Omega$. This plot allows to estimate how low frequencies can be discussed without losing accuracy for the approximate wave functions.
\begin{figure}[H]
	\centering
	\includegraphics[width=1\linewidth]{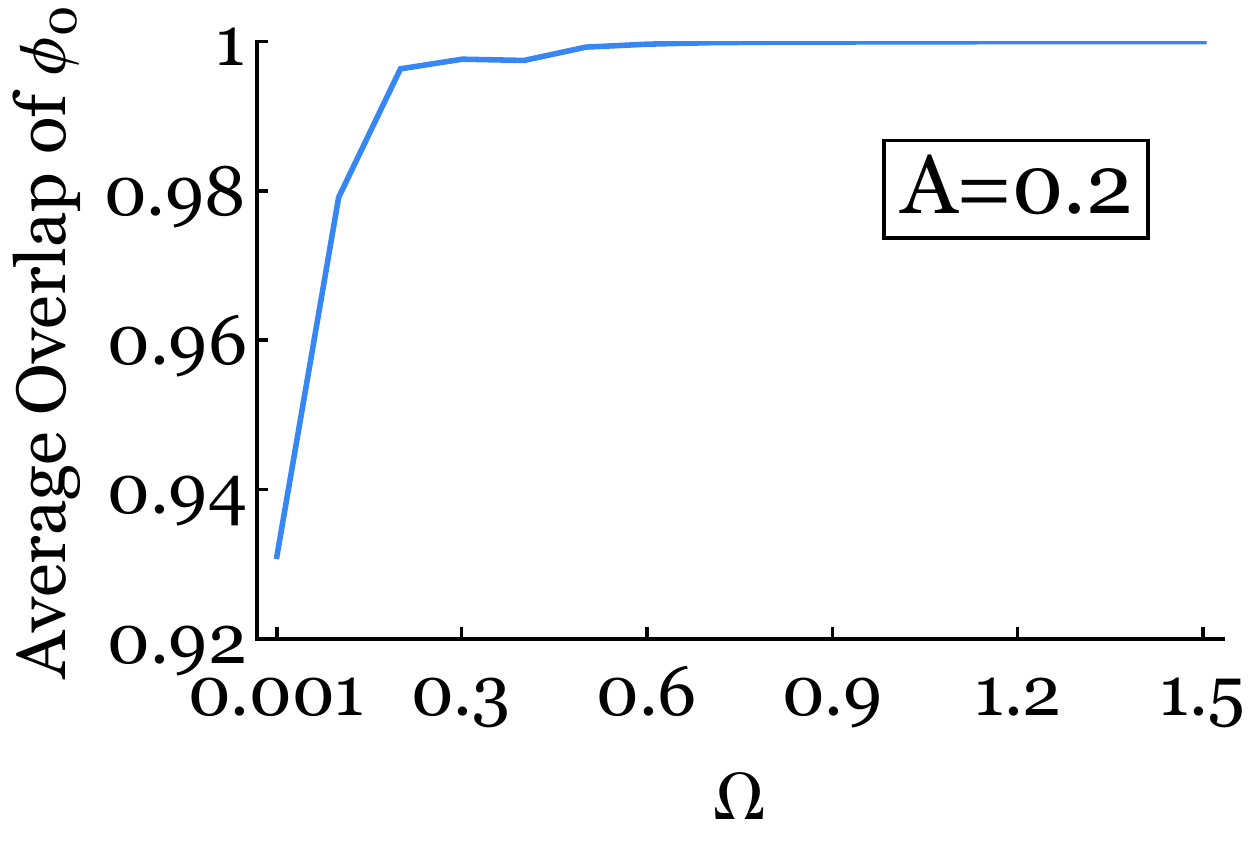}
	\caption{(Color online.) Overlap between our approximate result and the numeric solution of $\phi_0$ for Eq.\eqref{quasi-en-eq} with  $A=0.2$ solved for 16 Floquet modes (enough for numerical convergence) and averaged over $300$ points along the high-symmetry path $\Gamma\to K\to M\to \Gamma$. }
	\label{fig:OverLapAsFunctionFrequ}
\end{figure}

From the figure one may see that the approximation allows us to reach very small frequencies that are slightly smaller than the small driving strength $A$ before the approximation begins to deteriorate.

\begin{figure}
	\centering
	\includegraphics[width=1.0\linewidth]{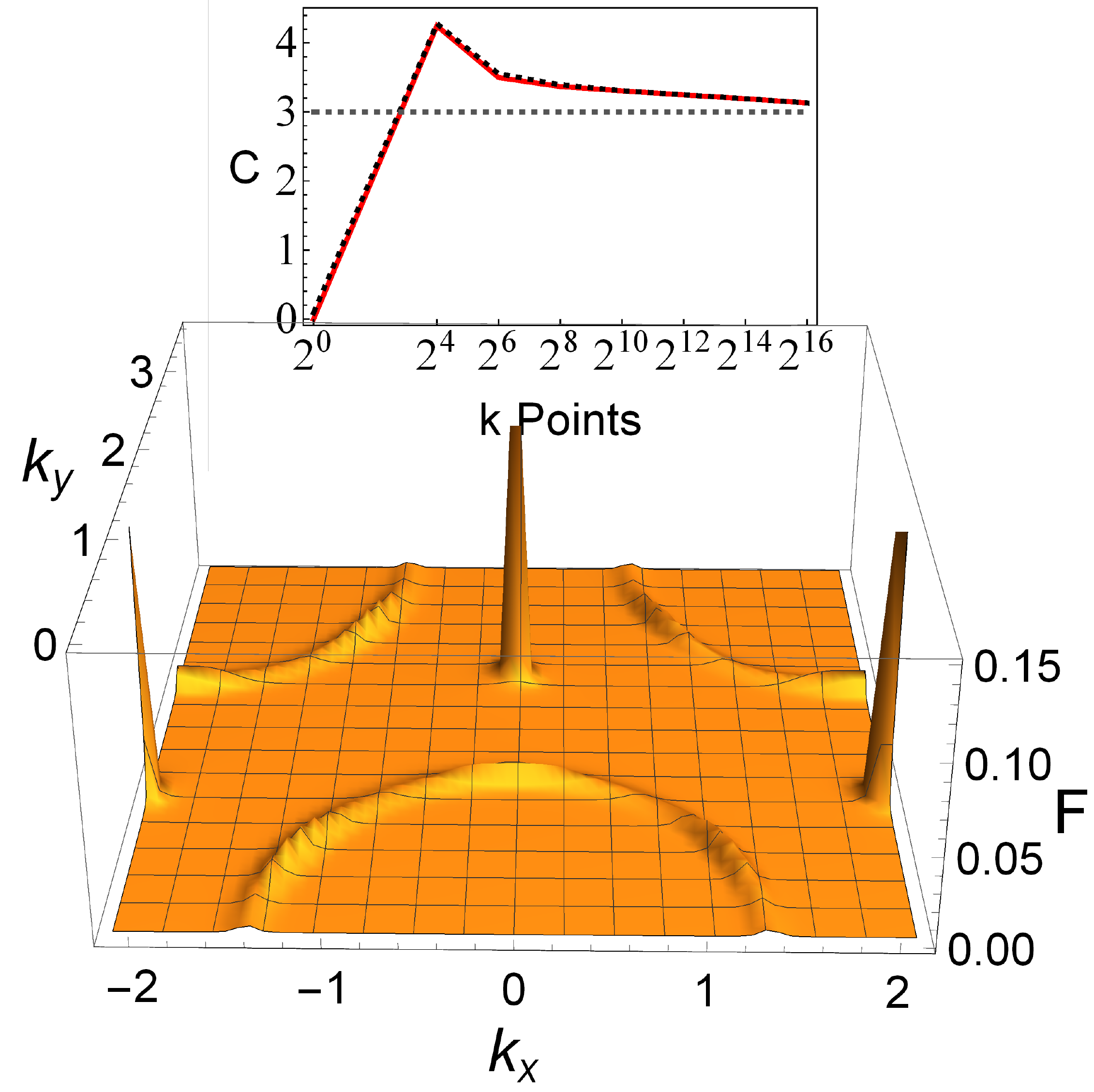}
	\caption{(Color online.) The Berry curvature $F$ for the lower band in the first Floquet zone with $\Omega=3.5$, $A=0.2$ and using $2^{12}$ points in $k$ space to obtain the plot. The plot shows the result obtained via our approximate method, which is visually indistinguishable from the result obtained by solving Eq.\eqref{quasi-en-eq} for three Floquet modes i.e. for $\phi_3,...,\phi_{-3}$. In the inset we see the numeric result for the Chern number plotted as a function of $k-$points. The black dashed curve hereby is the extended space result and the red curve the result obtained via our approximation. }
	\label{fig:berrycurvature}
\end{figure}

\textit{Berry curvature and Chern number.} We next focus on the lower of the two Floquet bands in the first Floquet zone. Appropriate care must be taken to ensure that $\phi_0$ is restricted this lower band. The proper procedure to fold the spectrum see \ref{fig:drawing-foldingstuff} is crucial. Regardless one may then use the standard approach described in Ref.[\onlinecite{doi:10.1143/JPSJ.74.1674}] to calculate the Berry curvature and Chern number,  seen in Fig.\ref{fig:berrycurvature}.

In the plots we see that both our approximation and the result obtained using the extended space method yield $C=3$. We may therefore state that the effective Floquet Hamiltonian here derived not just approximates quasi-energies well but is accurate enough to reliably predict topological properties, such as the Chern numbers. While $C$ is experimentally observable as stated by \cite{PhysRevA.91.043625}, it is not the topological invariant that defines the bulk-boundary correspondence in Floquet systems. For this, the appropriate invariants, which predict the number of edge states in a system with open boundary conditions, can be determined by following the procedure discussed by \cite{PhysRevX.3.031005}. Also it is worth mentioning that the slow convergence of $C$ is due to the small band-gap - indeed if we increase the band gap by using larger $A$ the quantity converges faster.

\textit{Relevance to experiments.} The theory presented here is defined in the limit of weak drives and low-frequencies. In the linear regime, the ultrafast response of the system is usually linearly proportional to the applied fluence. In principle, there is no lower-limit threshold for the amplitude strength, and experiments are limited by the detectable signal-to-noise ratio. Highly specialized high-sensitivity techniques can allow one to work with fluence as low as $\mu J/$cm$^2$. With respect to the drive frequency, current experimental techniques allow one to use frequencies as low as $0.5$ THz. In the case of far-infrared pulses, $65$~meV can be achieved. These experimentally-accessible frequencies allow one to study many materials in the low-frequency, weak drive regime. For example, monolayer transition metal dichalcogenide (TMDs), of interest for valleytronic applications, have typical band-gaps of the order of $1-2$~eV~\cite{xidong2015}. Additionally, spin-orbit effects lead to valence and conduction band splittings in the order of $100$~meV and $10$~meV respectively~\cite{Gui-Bin2013}. The drive frequencies required for the low-frequency regime are well in experimental reach. 

The weak-drive regime is also within reach. For example, the typical lattice constant for TMDs is $a_0 \sim 3$ \AA. Then, a laser fluence $f$ of the order of $\mu J/$cm$^2$, drive frequency $\sim 15.8$~THz ($\sim 65$~meV), and laser pulse duration $\tau = 0.1$~ps in a pump-probe setting gives $a_0 e A/\hbar = a_0 e E/(\hbar \Omega) = 0.25 < 1$, well within the reach of current experimental capabilities. Therefore, we expect that the method  introduced here to derive effective Floquet Hamiltonians will support the prediction and interpretation of experimentally-relevant Floquet systems. Furthermore, this work can form the foundation for further theoretical studies in the physics of low-frequency Floquet systems, particularly those involving interactions and coupling to auxiliary degrees of freedom, such as phonons and magnons. 

\textit{Acknowledgments.} We thank Edoardo Baldini for useful discussions. This work was supported by the NSF Materials Research Science and Engineering Center Grant No. DMR-1720595.

 \bibliography{literature}

\end{document}